\documentclass[11pt]{article}
\usepackage{geometry}                
\usepackage{graphicx}
\usepackage{amssymb}
\usepackage{amsmath}
\usepackage{amsthm}
\usepackage{epstopdf}
\usepackage{subfigure}
\usepackage{url}
\usepackage{xcolor}
\usepackage{gensymb}

\title{Building exploration with leeches \emph{Hirudo verbana}}
\author{Andrew Adamatzky and Georgios Ch. Sirakoulis\\
Unconventional Computing Centre and Bristol Robotics Lab,\\
University of the West of England, UK \\
Department of Electrical and Computer Engineering,\\
Democritus University of Thrace, Greece}

\begin{document}

\maketitle

\begin{abstract}
\noindent
Safe evacuation of people from building and outdoor environments, and search and rescue operations, always will remain actual in course of all socio-technological developments. Modern facilities offer a range of automated systems to guide residents towards emergency exists. The systems are assumed to be infallible. But what if they fail? How occupants not familiar with a building layout will be looking for exits in case of very limited visibility where tactile sensing is the only way to assess the environment? Analogous models of human behaviour, and socio-dynamics in general, are provided to be fruitful ways to explore alternative, or would-be scenarios. Crowd, or a single person, dynamics could be imitated using particle systems, reaction-diffusion chemical medium, electro-magnetic fields, or social insects. Each type of analogous model offer unique insights on behavioural patterns of natural systems in constrained geometries. In this particular paper we have chosen leeches to analyse patterns of exploration. Reasons are two-fold. First, when deprived from other stimuli leeches change their behavioural modes in an automated regime in response to mechanical stimulation. Therefore leeches can give us invaluable information on how human beings might behave under stress and limited visibility. Second, leeches are ideal blueprints of future soft-bodied rescue robots. Leeches have modular nervous circuitry  with a rich behavioral spectrum. Leeches are multi-functional, fault-tolerant with autonomous inter-segment coordination and adaptive decision-making. We aim to answer the question: how efficiently a real building can be explored and whether there any dependencies on the pathways of exploration and geometrical complexity of the building. In our case studies we use templates made on the floor plan of real building. 

\end{abstract}

\section{Introduction}

During the last decades there is a growing interest of scientists on the safety of people when gathering in indoor and outdoor environments like big buildings, sporting arenas, music halls, shopping centers, etc. The main issue in all these cases concerns the evacuation process in case of emergency and, more specifically, how it will be feasible to prevent accidents during it. It has been indicated that crowd safety and comfort not only depend on the design and the operation of the under study environment itself, but also relies on the behaviour of each crowd individual \cite{Helbing00}. From historical point of view, the early approach of motion prediction applied to large crowds of pedestrians was mainly based on the modelling of the crowd as a continuous homogeneous mass that behaves like a fluid flowing along corridors \cite{hughes2003flow}. Albeit, quest for more realistic and efficient in case of emergency, modern evacuation modelling approaches that will be able to reproduce phenomena like herding behaviour, clogging, arching near the exits, individualism, grouping and other behaviours, which are related with the process is still an on-going process. In the meantime, there are several works in literature \cite{duives2013state} dealing with these issues while indicating different approaches that try to envisage  major features of human behaviour, such as the decentralized crowd behavioral model \cite{reynolds2006big}, particle dynamics for grouping \cite{Brogan02}, social forces \cite{Helbing02}, data driven models \cite{Lerner07,Lee07,boukas14a}, etc. Moreover, in some of these models, pedestrians are ideally considered as homogeneous individuals, whereas in others, they are treated as heterogeneous groups with different features (e.g., gender, age, psychology). In general, crowd movement models can be categorized into \textit{top-down} or macroscopic and \textit{bottom-up} or microscopic ones \cite{tsiftsisreal}. Macroscopic models like lattice-gas \cite{li2008lattice} and fluid-dynamic \cite{Helbing02} focus on the total number of the members of the crowd ignoring possible differences on the individual behaviour. Microscopic models study the spatial and temporal behavior of each of the individuals and their interaction with the other members of the crowd. These are methods, where collective phenomena emerge from the complex interactions among individuals (self-organizing effects), thus describing pedestrian dynamics in a microscopic scale. Cellular Automata (CA) \cite{neumann66} models, agent-based model and social-force models belong to this category \cite{Vizzari13,Sirakoulisbook,Yu07,Nishinari06,Georgoudas06,Burstedde04}.

Analogous models of human behaviour, and socio-dynamics in general, are provided to be fruitful ways to explore alternative, or would-be scenarios. Crowd, or a single person, dynamics could be imitated using particle systems~\cite{chen2009controlling,heigeas2010physically}, reaction-diffusion chemical medium~\cite{adamatzky2005dynamics}, heat-transfer~\cite{liang2012cma}, electro-magnetic fields~\cite{su2014crowd}. {Nevertheless, most of the proposed models either agent-based or not try to mimic as close as possible human behaviour during evacuation and are considered to fulfill their promises based to the corresponding quantitative and qualitative results. The latest are considered a well known first approach on the effectiveness and the robustness of the models to reproduce the aforementioned phenomena synonymous to the prominent human crowd behavior during evacuation and in emergency situations. Towards this direction, scientists have used other species and living organisms as a fine substitute to both humans and corresponding models found in literature while trying to simulate evacuation in different environments. For example, Shiwakoti et al. \cite{Shiwakoti2011} in analogy to Burd et al., Couzin and Frank and Chowdhury et al. works on traffic modeling with different species of ants \cite{Burd2002,Couzin2003,Chowdhury2010} have used Argentine ants as a proxy for humans and studied their behaviour under panic to test different structural features to the panic escape in a chamber with fixed dimensions \cite{Shiwakoti2013}. In a similar manner, we have used slime mould to bio-mimic the human evacuation from a building and, furthermore, to develop a corresponding computational \textit{Physarum}-inspired crowd evacuation model based on CA by taking into account while mimicking the \textit{Physarum} foraging process, the food diffusion, the organism's growth, the creation of tubes for each organism, the selection of optimum tube for each human in correspondence to the crowd evacuation under study and finally, the movement of all humans at each time step towards near exit \cite{kalogeiton2015cellular}.}

{So it is clear that nowadays it is become more and more efficient to use living creatures as a real-world analog models and in a more generalized way} of spatially-extended computing and technological systems. Most famous instances of analog modelling with living substrates include laboratory experiments with real ants on improved collective performance in distributed tasks, decision making and robotics~\cite{langridge2008experienced, deneubourg1991dynamics}; design and implementation of logical gates with soldier crabs~\cite{gunji2011adaptive}; development and manufacturing of sensors and computing circuits with slime mould \emph{Physarum polycephalum}~\cite{adamatzky2010physarum}. 

In present paper we employ leeches to study explorations of building in scenarios with limited visibility and sensing. A leech \emph{Hirudo medicinal} and its South European analog \emph{Hirudo verbana} are amongst most common living creatures explored in laboratory conditions. Leeches are ideal inspirations for amphibious  soft or flexible search-and-rescue robots capable for reaching spaces not accessible by other devices amphibious~\cite{crespi2005swimming, crespi2004amphibious, yang2007preliminary, crespi2005amphibot, yang2008body, yu2009amphibious}. The reasons are following. 

The leeches' neural networks are simple yet efficient, they are equivalent in their computational power to basic perceptrons~\cite{lockery1993computational}. Notably, a nervous system of leech became a test bed for modelling locomotion control~\cite{KristanJr1977191, lockery1993computational, campos2007temporal, gaudry2010feeding, crisp2012mechanisms, kristan2005neuronal, friesen1993mechanisms, brodfuehrer1993effect, lockery1993lower}, modulating behaviour of neuro-mediators~\cite{zaccardi2004sensitization, alkatout2007serotonin, gerry2012serotonin},  developmental processes in complex neuronal circuits~\cite{kristan2000development},  and mathematical and computers models of circuits responsible for regular pattern generations~\cite{taylor2000model, zheng2007systems, pearce1988model, buono2004mathematical}.  

A leech has exactly thirty two segments. A single segment of the leech's body contains isolated ganglion. It is capable for exhibiting swimming activity even when the segment has been neurally isolated from the rest of the leech's body~\cite{KristanJr1977191}. 

A spectrum of leeches' behaviour traits is extensively classified~\cite{dickinson1984feeding}. The following pattern is reported by Dickinson and Lent~\cite{dickinson1984feeding}.   A leech positions itself at the water surface in resting state. The leech swims towards the source of a mechanical or optical stimulation. The leech stops swimming when comes into contact with any geometrical surface. Then, the leech explores the surface by crawling. When a leech finds a warm (37-40$^o$C) region the leech bites. There is also useful feature of context-modulate behaviour: a leech can respond to constant sensorial inputs with variable motor outputs~\cite{brodfuehrer2001identified}.

{Finally, b}ehaviour of leeches in uniform spaces is well analyse. There leeches wander around uniformly and no preferential direction or location have been observed~\cite{garcia2005statistics}.
Results on leeches bevavior in complex geometries are scarce or non-existent. It is know that leeches show positive stigmotaxis and therefore crawl under logs for hiding or body cavities feeding. Till our recent paper \cite{Adamatzky201528} no result were on how geometrical constraints of a space shape leech's behavioural patterns. In \cite{Adamatzky201528} we found that a leech switches into exploration mode when it encounters a mechanical obstacle and a probability of returning from exploratory mode to crawling mode decreases proportionally to a distance from last mechanical obstacle. Therefore, when placed in a corridor with a raw of rooms on one side the leech explores rooms near end of the corridor with higher probability and rooms near the centre of the corridor with lower probability. Based on the results of our laboratory experiments we formalised behaviour of a leech in terms of probabilistic finite state machines with binary inputs. {In the present paper we decided to go further and to study how leeches explore geometrical constrained areas with templates matching part of real buildings. The rationals for this study are as follows. What if a person tries to move towards an exit in total darkness and silence and absence of smell, by relying just on their tactile stimuli and having no previous knowledge of the building layout. How does pattern of exploration develop? Will the person explore the whole building? Or will stack in some particular parts? It should be further defined that the usage of such a scenario, i.e. where the human has limited senses and/or information received by the under study environment enables the possibility of bio-mimicking the human behaviour by a less complex biological entity with diminished behavioral abilities compared to humans like leeches. In such a way, the provided study and the corresponding results of the leeches under panic (the term ''panic" should be also defined in a different base compared to the one arisen by the sociological science and referring to not rational behavioral due to emotional stress) sound promising without studying the physical and behavioral similarities and dissimilarities among leeches and humans in one-to-one basis.}

\section{Methods}

We used three weeks old leeches \emph{Hirudo verbana} obtained from Biopharm Leeches (Hendy, Carmarthenshire SA4 0X, UK). Leeches varied in size from 10 to 20~mm length in elongated state and 1-2~mm width. Leeches awaiting experiments were kept in securely covered, yet with air access, glass containers in a dechlorinated water away from direct sunlight. As per recommendation~\cite{taneja2011national} leeches were kept in a cool, c. 15$\degree$C, environment to lessen their needs for feeding and to enhance their performance in exploration of experimental templates. The water was refreshed every other day. When moving leeches between storage containers and experimental templates we used non-serrated forceps. 

In experiments we used template of the first floor of building B of the Electrical and Computer Engineering Department (ECE) of the Democritus University of Thrace (DUTh). The floor plan of the space is accurate and consistent with the real dimensions of the university building. This particular office layout was chosen because a range of experiments undertaken and computer models developed for this particular layout of rooms~\cite{kalogeiton2014hey,kalogeiton2015cellular}. {By using the same template we make experiments with leeches compatible with our previous results and data obtained. The scenario adopted here considers a human randomly located in the aforementioned space. Please consider that no previous knowledge of the under exploration environment is provided to her/him while there is no light and sound and hence the human is not able to see or hear anything. By assuming that she/he is not able to use any of her/his senses, e.g. sense of touch, and is equipped only with a torch. The same constrains correspond in analogy to the leeches that will be used for the exploration of the template.}

\begin{figure}[!tbp]
\centering
\includegraphics[width=0.7\textwidth]{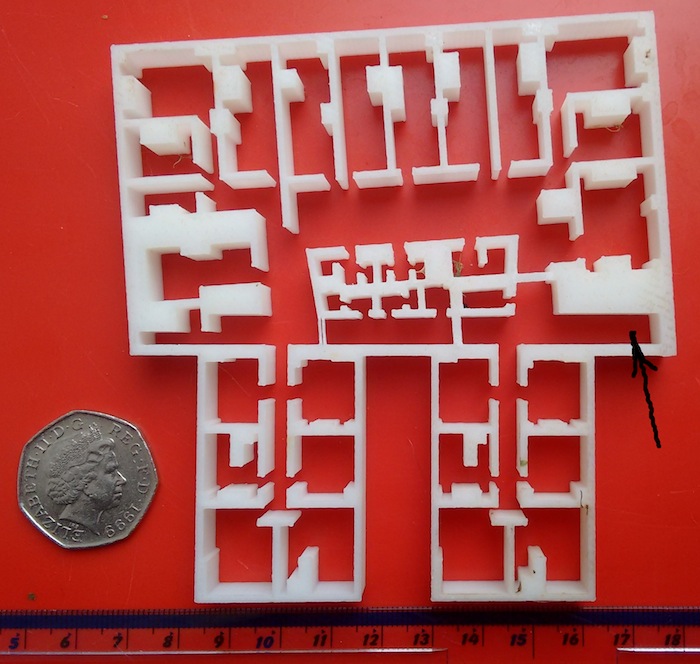}
\caption{Photo of geometrical template used. Initial position of leech is shown by arrow.}
\label{fig:template}
\end{figure}

The template was printed from polylactic acid thermoelastic polyester (Fig.~\ref{fig:template}) with the maximum dimensions 11 cm by 10 cm and wall height 1 cm. The layout contains 24 rooms, each room has opening to corridors.  Experimental template was cleaned to remove any substances with odour or state that might affect behaviour of leeches. The template was filled with dechlorinated water, depth 9~mm; the leeches  were able to swim if they wanted {too}. The water was renewed to prevent metabolites and ions released by leeches to affect behaviour of their successors. The template was illuminated by LED lamp and the illumination level at the bottom of the template was 37 LUX. No sharp gradients of optical, chemical or electrical stimuli were allowed; the only stimulation occurred was mechanical one when leeches come into contact with walls of the template. Experiments were conducted in a room temperature of 20$\degree$C. We conducted 20 experiments using 20 leeches; each leech was used once.

The experiments were recorded on Coolpix P90 digital camera, 640$\times$480 pixels frame size and 25 frames per second speed. Each video was recorded for c. 30~min or till leech escapes, whatever happens early. The videos were analysed by in-house software written in Processing, as follows. For every second of video we extracted coordinates of pixels with colour values less than 30-50 in RGB mode (exact threshold was adjusted for each video). Such pixel represented body of the leach. Their coordinates with time tags were stored for further analyses. Configurations of leeches exploring the template were converted to overlay images with colours  as follows. For any trial/video a duration of recording is normalised to the interval [0,1] and then mapped to a colour scale $(00B)$ $\rightarrow$  $(0BG)$ $\rightarrow$ $(RG0)$ $\rightarrow$ $(R00)$, i.e. the blue pixel represent leech at the beginning of experiment and red pixel at the end of experiment. 

\begin{figure}[!tbp]
\centering
\includegraphics[width=0.7\textwidth]{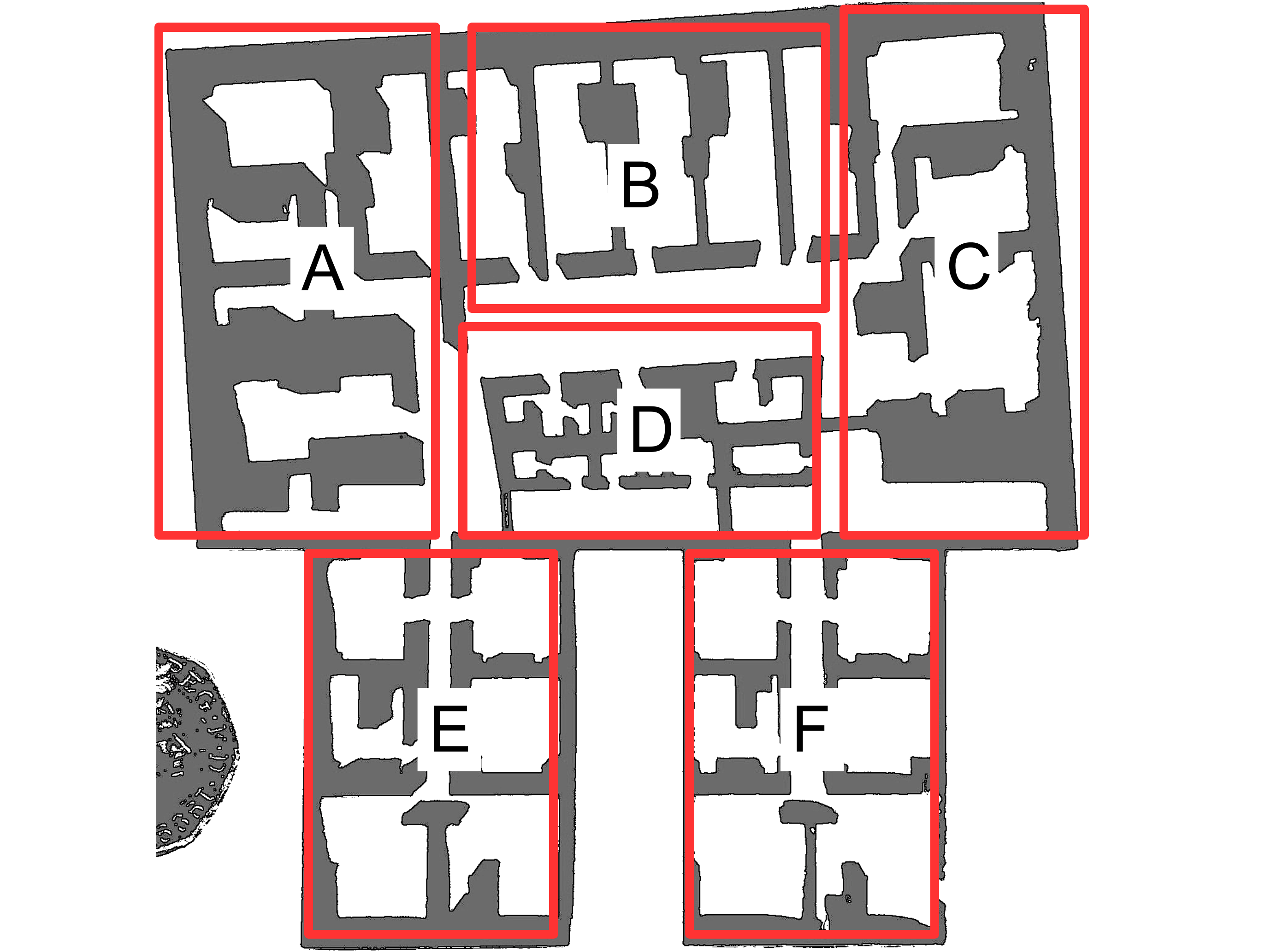}
\caption{Separation of template into several domains.}
\label{fig:boxes}
\end{figure}

We separate the layout into six domains (Fig.~\ref{fig:boxes}) which will be referred to in further analysis and discussions.

\section{Results}

\begin{figure}[!tbp]
\centering
\subfigure[1 min]{\includegraphics[width=0.3\textwidth]{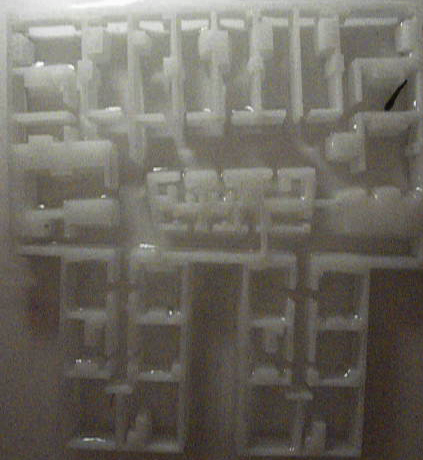}}
\subfigure[1 min 48 sec]{\includegraphics[width=0.3\textwidth]{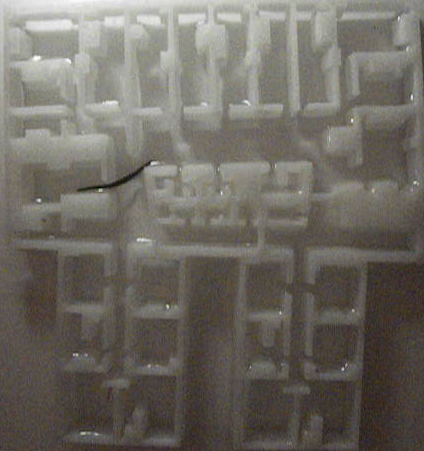}}
\subfigure[2 min 39 sec]{\includegraphics[width=0.3\textwidth]{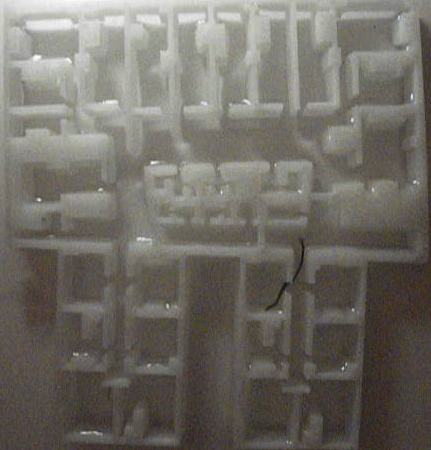}}
\subfigure[5 min 14 sec]{\includegraphics[width=0.3\textwidth]{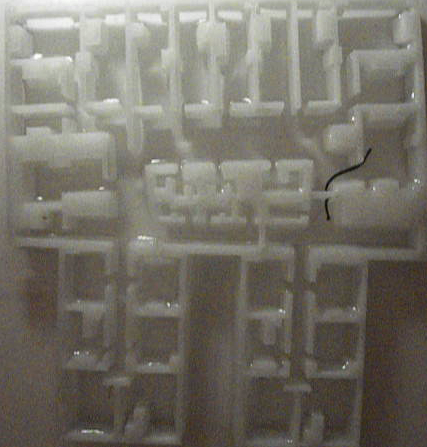}}
\subfigure[13 min 30 sec]{\includegraphics[width=0.3\textwidth]{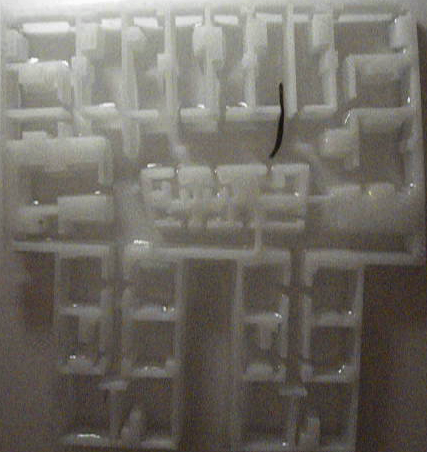}}
\subfigure[23 min 26 sec]{\includegraphics[width=0.3\textwidth]{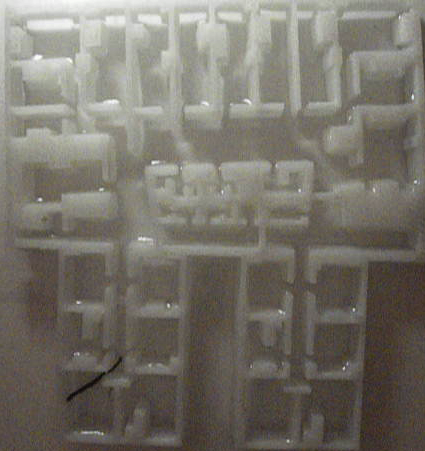}}
\caption{Snapshots of a leech exploring template.}
\label{fig:snapshots}
\end{figure}

Snapshots of an experiment are shown in Fig.~\ref{fig:snapshots}. The leech was placed {by random selection} in the right end of corridor of domain $C$, as indicated by arrow in Fig.~\ref{fig:template}. In first minute, leech propagated to room right cornet room in domain $C$ (Fig.~\ref{fig:snapshots}(a)). It then explores the domain $C$ and leaves for domain $D$, which is passed almost without exploration (Fig.~\ref{fig:snapshots}(b)). In roughly two and half minutes the leech enters corridor in domain $F$ and immediately proceeds to the first room near the entrance of the domain (Fig.~\ref{fig:snapshots}(c)). After exploring rooms in domain $F$ the leech returned to domain $C$ (Fig.~\ref{fig:snapshots}(d)) and then swim to domain $B$ (Fig.~\ref{fig:snapshots}(e)). Domain $E$ is getting explored after twenty minutes of the experiment (Fig.~\ref{fig:snapshots}(f)).

\begin{figure}[!tbp]
\centering
\subfigure[]{\includegraphics[width=0.3\textwidth]{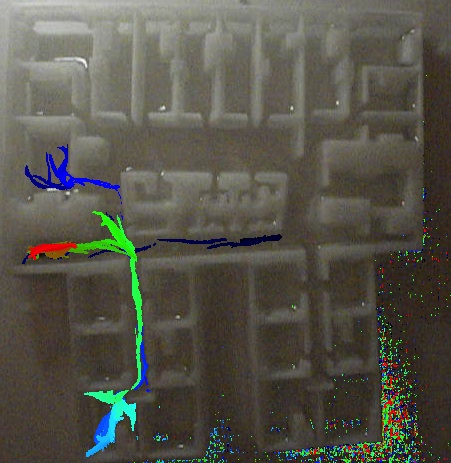}}
\subfigure[]{\includegraphics[width=0.3\textwidth]{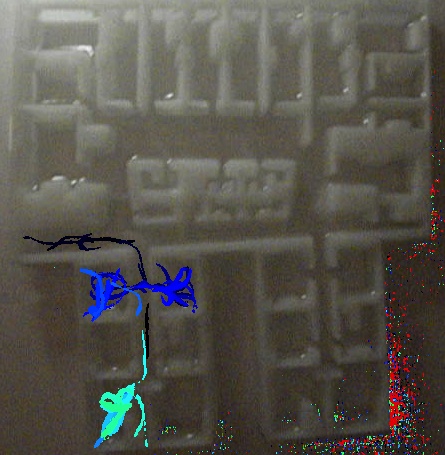}}
\subfigure[]{\includegraphics[width=0.3\textwidth]{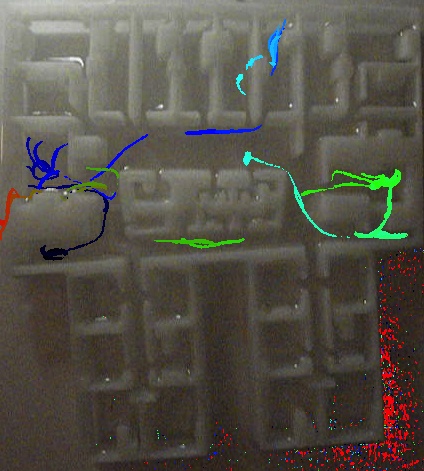}}
\subfigure[]{\includegraphics[width=0.3\textwidth]{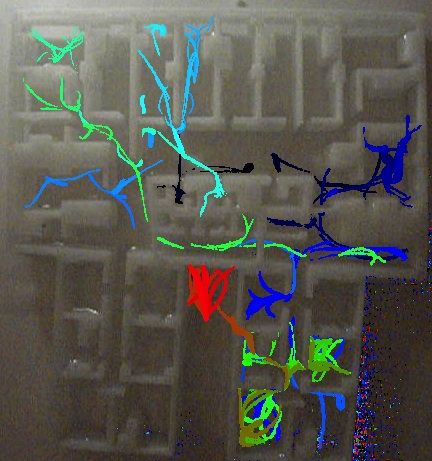}}
\subfigure[]{\includegraphics[width=0.3\textwidth]{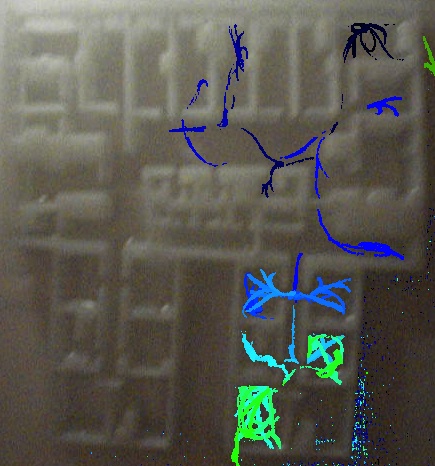}}
\subfigure[]{\includegraphics[width=0.3\textwidth]{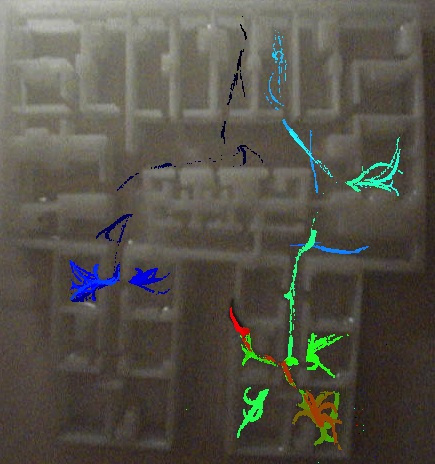}}
\subfigure[]{\includegraphics[width=0.3\textwidth]{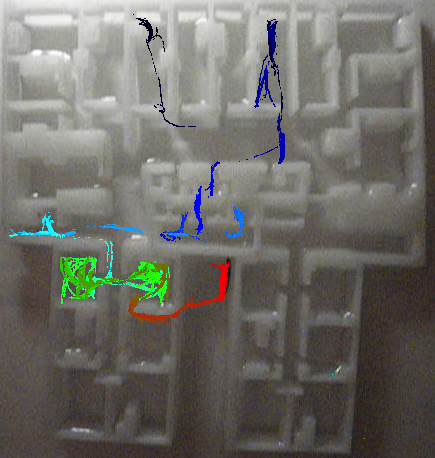}}
\subfigure[]{\includegraphics[width=0.3\textwidth]{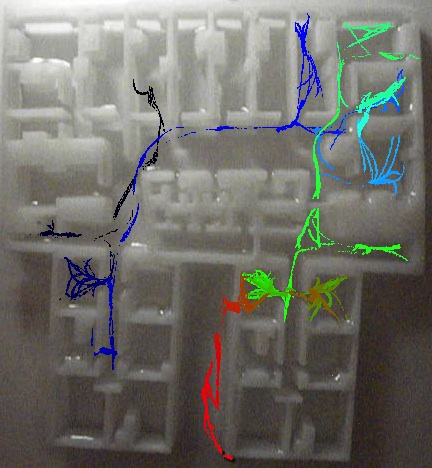}}
\subfigure[]{\includegraphics[width=0.3\textwidth]{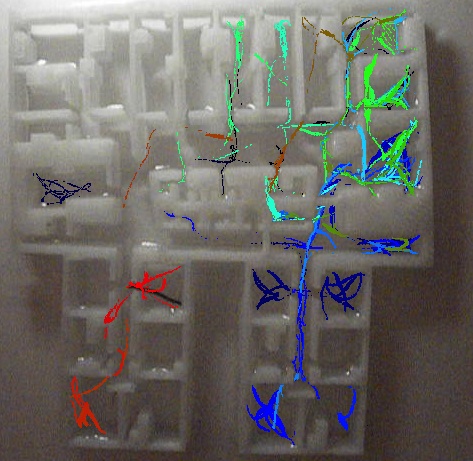}}
\caption{Time snapshots of selected experiments.}
\label{fig:timeoverlay}
\end{figure}

Time to color coded overlays of leeches in representative experiments are shown in Fig.~\ref{fig:timeoverlay}. In experiments shown in Figs.~\ref{fig:timeoverlay}(a) and (b), leeches explore mainly domain $E$, and they spent most of their time in domains $A$ to $D$ in experiment Fig.~\ref{fig:timeoverlay}(c). A leech visits almost all domains by $E$ in experiment Fig.~\ref{fig:timeoverlay}(d). A sequence of domains visited, as seen in color coding is $C$, $F$, $D$, $B$, $A$, $D$, $F$. In this particular the leech escaped outside template from the domain $F$. Other illustrations show leeches predominantly visiting $F$ (Figs.~\ref{fig:timeoverlay}(e) and (f)), $E$ (Fig.~\ref{fig:timeoverlay}(g)) and both $E$ and $F$ (Figs.~\ref{fig:timeoverlay}(h) and (i)).

\begin{figure}[!tbp]
\centering
\includegraphics[width=0.8\textwidth]{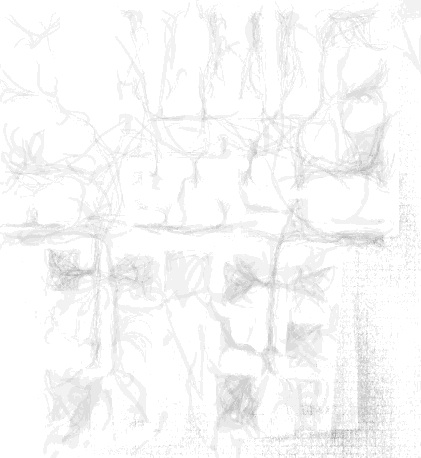}
\caption{Frequency matrix normalized to gray level: 0 is white and 1 is black.}
\label{frequency}
\end{figure}

\begin{figure}[!tbp]
\centering
\subfigure[$\theta=0$]{\includegraphics[scale=0.4]{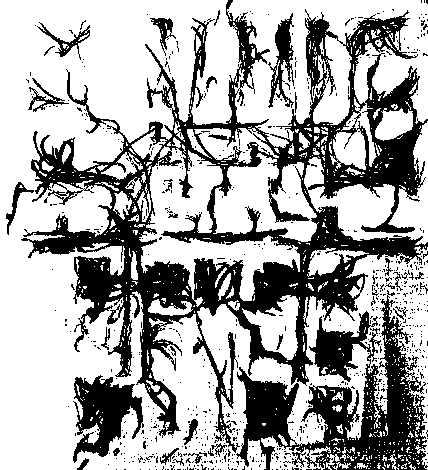}}
\subfigure[$\theta=0.05$]{\includegraphics[scale=0.4]{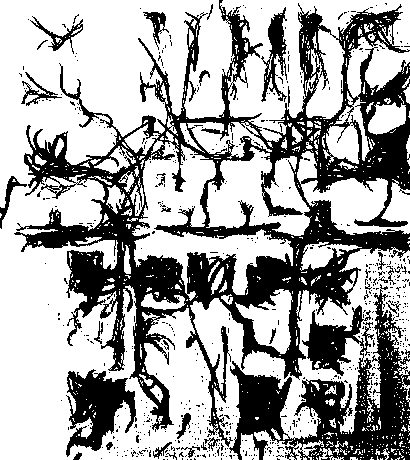}}
\subfigure[$\theta=0.1$]{\includegraphics[scale=0.4]{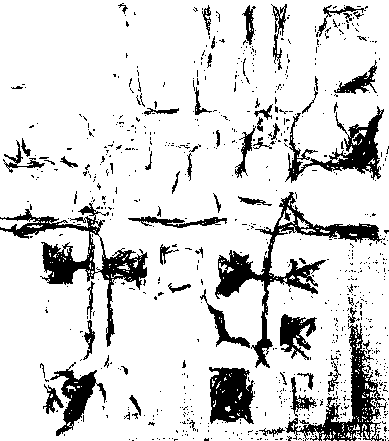}}
\subfigure[$\theta=0.15$]{\includegraphics[scale=0.4]{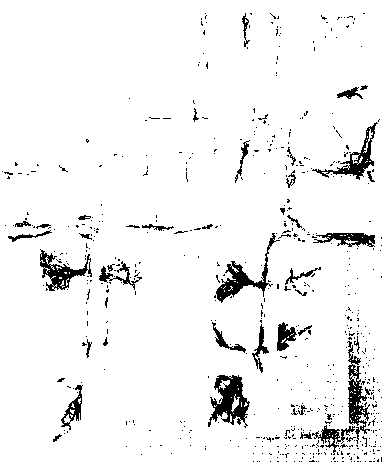}}
\caption{Threshold frequency matrix normalized to gray level. Only non-zero entries exceed threshold $\theta$ are shown by black pixels. Values of $\theta$ are shown in the captions to sub-figures.}
\label{thresholdfrequency}
\end{figure}

We approximate a frequency of a leech to be in site $x$ of a template as a number of experiments leech visited site $x$ normalized by a total number of sites visited by leeches in all experiments. The frequency of visits might also reflect time a leech spends in the domain, in cases when leech does not remain still. The frequency of visits field of the experimental template is shown in Fig.~\ref{frequency}. Frequency cut off for various levels of thresholds are presented in Fig.~\ref{thresholdfrequency}. From the frequency matrix (Fig.~\ref{frequency}) we can calculate frequencies of leeches visiting domains: 
$f(A)=0.09$, 
$f(B)=0.11$,
$f(C)=0.17$,  
$f(D)=0.11$,
$f(E)=0.23$,
$f(F)=0.30$. 
Thus we have the following hierarchy of leeches' preferences in visiting particular domains: 

\begin{equation}
f(F) > f(E) > f(C) > f(A) > \{ f(B), f(D) \} .
\label{fhierarchy}
\end{equation}

We see that leeches prefer to visit domain $F$ and visiting more sites inside the domain $F$ than in other domains of the template. Domain $E$ is second most visited part of the template. Further preferences to go domain $C$. Domains $A$, $B$ and $D$ are less visited part of the template, or less explored domains. 

\begin{figure}[!tbp]
\centering
\includegraphics[width=0.8\textwidth]{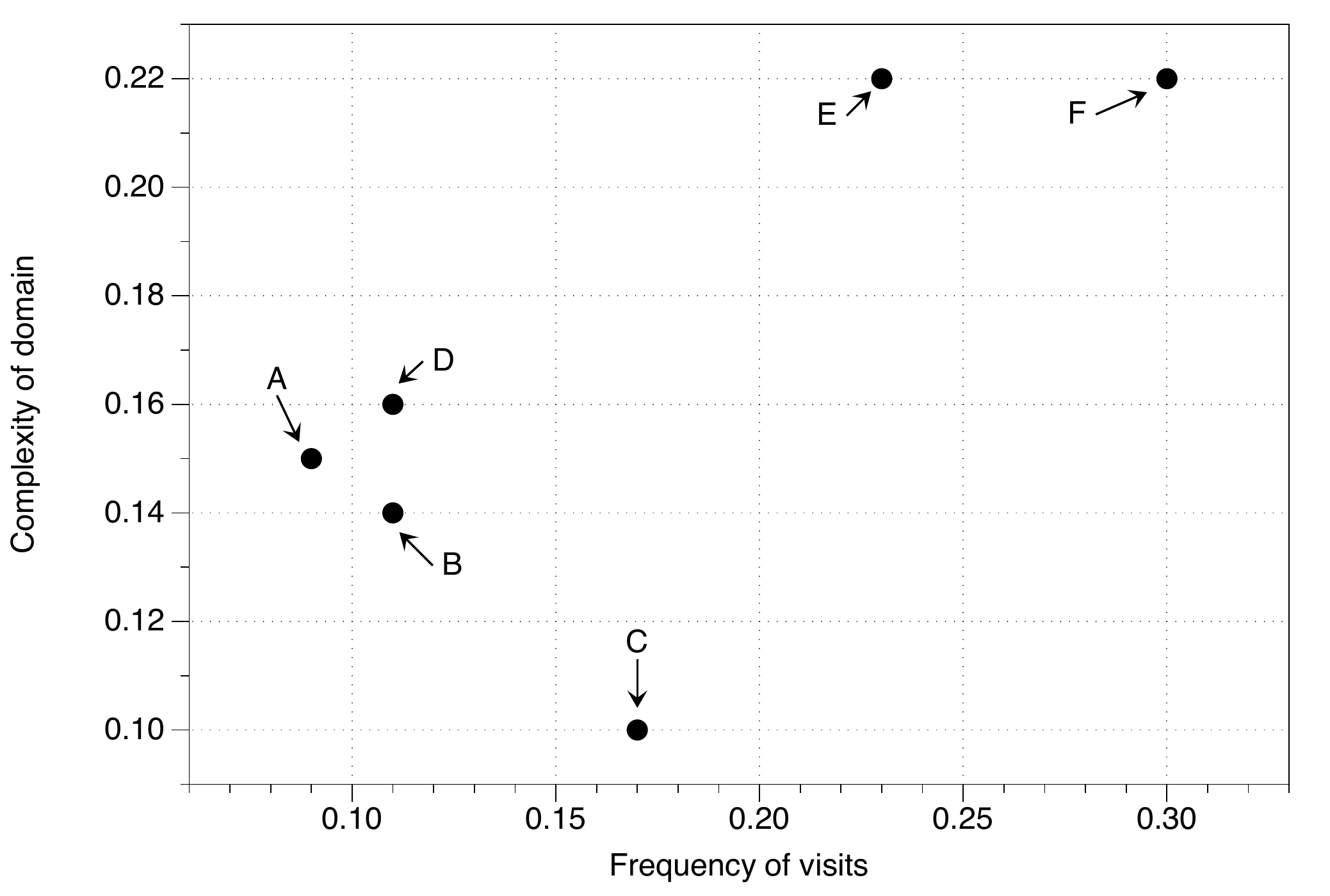}
\caption{Plot of frequency of leeches' visits to domains $f(\cdot)$ and complexity of the domains $c(\cdot)$.}
\label{plot}
\end{figure}

Is the hierarchy (\ref{fhierarchy}) determined by complexity of the template? 
We calculate complexity of a domain as a number of corners in the domain normalised by total number of corners in the template. {The idea is consistent with the findings of the many works reported in literature regarding the role of corners in indoor environments during evacuation under emergency.} Thus, we get the following list of complexities: 
$c(A)=0.15$,
$c(B)=0.14$,
$c(C)=0.10$,
$c(D)=0.16$,
$c(E)=0.22$,
$c(F)=0.22$. 
So in terms of complexity we have the following hierarchy:
\begin{equation}
\{ c(E), c(F) \} > c(D) > \{ c(A), c(B) \} > c(C) .
\label{complexity}
\end{equation}
As we can see in the frequency of visits versus complexity of domains plot Fig.~\ref{plot}, there three clusters of domains: 
\begin{enumerate}
\item moderate complexity and low frequency of visits: domains $A$, $B$, $D$
\item low complexity and moderate frequency of visits: domain $C$ 
\item high complexity and high frequency of visits: domains $E$ and $F$.
\end{enumerate}
Thus, we can state that \emph{a leech spends more time exploring floor layout templates with high complexity measured via number of corners}.

With regards to differences in visits frequencies between domains with the complexity, our explanations are as follows. Frequency $f(F)$ is 1.3 times higher frequency than $f(E)$ because the domain $F$ is closer to initial position of a leech (shown by arrow in Fig.~\ref{fig:template}). Thus a leech either turns into the corridor leading to rooms in the domain $F$, and then spends substantial amount of tie `bouncing' there between rooms, or propagate to the end of main corridor and then turns into the domain $E$.

\section{Discussion}

We used living leeches in physical analog modelling of buildings exploration in scenario of limited visibility and no guidance. We studied patterns of leeches behaviour during 'unguided' exploration of geometrically constrained space where no gradients of attractant or repellents are present. The leeches were only relying on mechanical stimulation by walls of the template.  We found that complexity of the building, even such simple one as measured via a number of corners in each part of the building, could provide a reliable estimate on how long time the leech will spend in the domain. Parts of the building with highest complexity are explored with higher degree of possibility. {Having also in mind that part $D$ of the examined floor corresponds to the ''core" of the building with toilets, showers, communication rooms and store houses, i.e. rather small rooms, this is also rational in terms of evacuation decisions and strategy.} 

\begin{figure}[!tbp]
\centering
\subfigure[]{\includegraphics[scale=0.45]{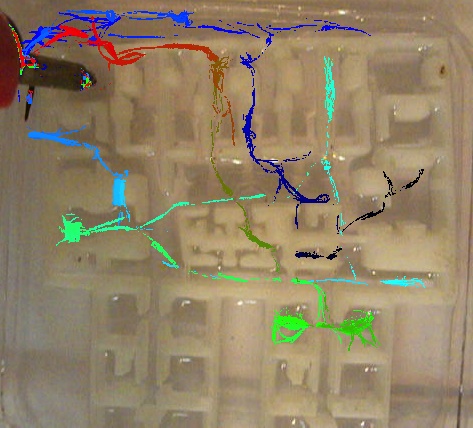}}
\subfigure[]{\includegraphics[scale=0.45]{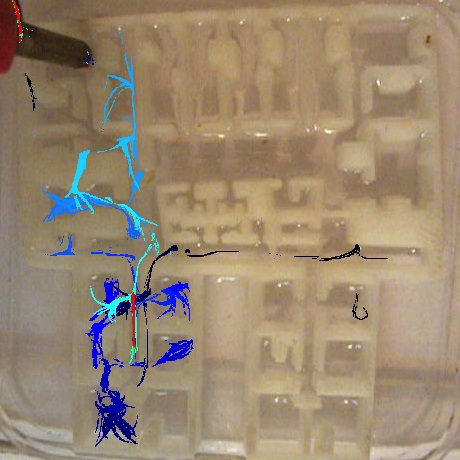}}
\subfigure[]{\includegraphics[scale=0.45]{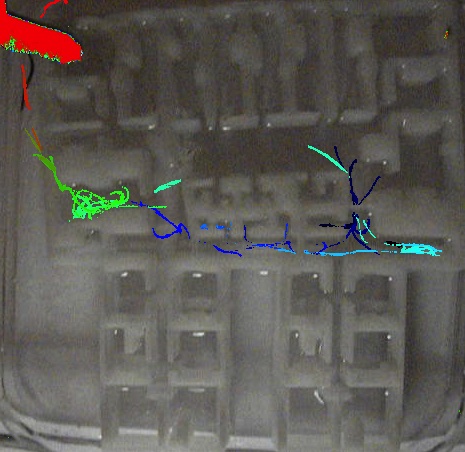}}
\caption{Overlay of leeches movement in template with a source of thermal gradients.}
\label{thermal}
\end{figure}

Future studies could be concerned with imitating search and rescue missions with leeches. We can represent targets, a leech search, by sources of physical stimuli which attract leeches. Potential attracts could include mechanical waves in water, temperature and chemical attraction. We have conducted scoping with leeches behaviour in thermal gradients. To form the gradient we immersed, by 5~mm, a tip of a soldering iron, heated to 70$\degree$C, in the water inside top corner room, in the domain $A$. Leeches were placed in the original position on the right side of the corridor (shown by arrow in Fig.~\ref{fig:template}). 

In five (5) of eleven (11) experiments a leech escaped from the template, in three (3) experiments the leeches moved into the domains $E$ and $F$. However in three (3) experiments the leeches moved towards the source of thermal stimulation. For example, in experiment illustrated in Fig.~\ref{thermal}(a), a leech propagated in domain $D$, explored domain $B$, briefly explored $F$ and then moved to $A$. In experiment shown in Fig.~\ref{thermal}(b) the leech propagated to domain $E$ and spent a substantial amount of time exploring rooms there, then moved to domain $A$. In the experiment shown in Fig.~\ref{thermal}(c) the leech first traveled back and forth between the domains $F$ and $C$ and then moved to domain $A$. 

Future experiments could deal with rigorous analysis of thermal and vibrational gradients affecting the leeches propagation inside templates, and physical analog modelling of leeches templates or models of multi-storey buildings.

{Finally, we do intend to run similar scenarios with humans for the exploration and evacuation of the real building under study. It is clear that in correspondence to the evacuation management and the human behaviour, as stated by Omer and Alon \cite{Omer1994}, two deeply rooted misconceptions should be further considered; the first one, that was termed \textit{abnormalcy bias}, consists in underestimating the ability of people to function adequately in the face of disaster while the second, known as \textit{normalcy bias}, consists in underestimating the probability of disaster, or the disruption involved in it (see Drabek, 1986, for a review on the prevalence of these biases and their deleterious influence in disaster management). In such a sense the differences and the physical disimilarities between the human and the biological entities like leeches should be further investigated even in simple almost ground truth scenarios like the ones introduced in this paper.}

\section{Supplementary materials}
Videos of leeches:
\url{https://www.youtube.com/playlist?list=PLw-0L7RLQBmb8hn-xINBtAxhdN6Cq6VdU}

\bibliographystyle{plain}
\bibliography{bibliographyleeches}

\begin{thebibliography}{10}

\bibitem{adamatzky2005dynamics}
Andrew Adamatzky.
\newblock {\em Dynamics of crowd-minds: patterns of irrationality in emotions,
  beliefs, and actions}, volume~54.
\newblock World Scientific, 2005.

\bibitem{adamatzky2010physarum}
Andrew Adamatzky.
\newblock {\em Physarum machines: computers from slime mould}, volume~74.
\newblock World Scientific, 2010.

\bibitem{Adamatzky201528}
Andrew Adamatzky.
\newblock On exploration of geometrically constrained space by medicinal
  leeches hirudo verbana.
\newblock {\em Biosystems}, 130(0):28 -- 36, 2015.

\bibitem{alkatout2007serotonin}
Bilal~A Alkatout, Nicole~M Marvin, and Kevin~M Crisp.
\newblock Serotonin delays habituation of leech swim response to touch.
\newblock {\em Behavioural brain research}, 182(1):145--149, 2007.

\bibitem{boukas14a}
Evangelos Boukas, Luca Crociani, Sara Manzoni, Giuseppe Vizzari, Antonios
  Gasteratos, and Georgios~Ch. Sirakoulis.
\newblock An intelligent tool for the automated evaluation of pedestrian
  simulation.
\newblock In Aristidis Likas, Konstantinos Blekas, and Dimitris Kalles,
  editors, {\em Artificial Intelligence: Methods and Applications}, volume 8445
  of {\em Lecture Notes in Computer Science}, pages 136--149. Springer
  International Publishing, 2014.

\bibitem{brodfuehrer1993effect}
Peter~D Brodfuehrer, Andreas~M Kogelnik, W~Otto~Friesen, and Avis~H Cohen.
\newblock Effect of the tail ganglion on swimming activity in the leech.
\newblock {\em Behavioral and neural biology}, 59(2):162--166, 1993.

\bibitem{brodfuehrer2001identified}
Peter~D Brodfuehrer and Maria Stella~E Thorogood.
\newblock Identified neurons and leech swimming behavior.
\newblock {\em Progress in neurobiology}, 63(4):371--381, 2001.

\bibitem{Brogan02}
David~C Brogan and Jessica~K Hodgins.
\newblock Simulation level of detail for multiagent control.
\newblock In {\em Proceedings of the first international joint conference on
  Autonomous agents and multiagent systems: part 1}, pages 199--206. ACM, 2002.

\bibitem{buono2004mathematical}
Pietro-Luciano Buono and A~Palacios.
\newblock A mathematical model of motorneuron dynamics in the heartbeat of the
  leech.
\newblock {\em Physica D: Nonlinear Phenomena}, 188(3):292--313, 2004.

\bibitem{Burd2002}
Martin Burd, Debbie Archer, Nuvan Aranwela, and David J.~Stradling.
\newblock Traffic dynamics of the leaf‐cutting ant, atta cephalotes. .
\newblock {\em The American Naturalist}, 159(3):pp. 283--293, 2002.

\bibitem{Burstedde04}
Carsten Burstedde, Kai Klauck, Andreas Schadschneider, and Johannes Zittartz.
\newblock Simulation of pedestrian dynamics using a two-dimensional cellular
  automaton.
\newblock {\em Physica A: Statistical Mechanics and its Applications},
  295(3):507--525, 2001.

\bibitem{campos2007temporal}
Doris Campos, Carlos Aguirre, Eduardo Serrano, Francisco
  de~Borja~Rodr{\'\i}guez, Gonzalo~G de~Polavieja, and Pablo Varona.
\newblock Temporal structure in the bursting activity of the leech heartbeat
  cpg neurons.
\newblock {\em Neurocomputing}, 70(10):1792--1796, 2007.

\bibitem{chen2009controlling}
Ying-ping Chen and Ying-yin Lin.
\newblock Controlling the movement of crowds in computer graphics by using the
  mechanism of particle swarm optimization.
\newblock {\em Applied Soft Computing}, 9(3):1170--1176, 2009.

\bibitem{Chowdhury2010}
Debashish Chowdhury, Katsuhiro Nishinari, and Andreas Schadschneider.
\newblock Ca modeling of ant-traffic on trails.
\newblock In Jiri Kroc, Peter~M.A. Sloot, and Alfons~G. Hoekstra, editors, {\em
  Simulating Complex Systems by Cellular Automata}, Understanding Complex
  Systems, pages 275--300. Springer Berlin Heidelberg, 2010.

\bibitem{Couzin2003}
I.~D. Couzin and N.~R. Franks.
\newblock Self-organized lane formation and optimized traffic flow in army
  ants.
\newblock {\em Proceedings of the Royal Society of London B: Biological
  Sciences}, 270(1511):139--146, 2003.

\bibitem{crespi2004amphibious}
Alessandro Crespi, Andr{\'e} Badertscher, Andr{\'e} Guignard, and Auke~Jan
  Ijspeert.
\newblock An amphibious robot capable of snake and lamprey-like locomotion.
\newblock In {\em Proceedings of the 35th international symposium on robotics
  (ISR 2004)}, number BIOROB-CONF-2004-003, 2004.

\bibitem{crespi2005amphibot}
Alessandro Crespi, Andr{\'e} Badertscher, Andr{\'e} Guignard, and Auke~Jan
  Ijspeert.
\newblock Amphibot i: an amphibious snake-like robot.
\newblock {\em Robotics and Autonomous Systems}, 50(4):163--175, 2005.

\bibitem{crespi2005swimming}
Alessandro Crespi, Andr{\'e} Badertscher, Andr{\'e} Guignard, and Auke~Jan
  Ijspeert.
\newblock Swimming and crawling with an amphibious snake robot.
\newblock In {\em Robotics and Automation, 2005. ICRA 2005. Proceedings of the
  2005 IEEE International Conference on}, pages 3024--3028. IEEE, 2005.

\bibitem{crisp2012mechanisms}
Kevin~M Crisp, Brian~R Gallagher, and Karen~A Mesce.
\newblock Mechanisms contributing to the dopamine induction of crawl-like
  bursting in leech motoneurons.
\newblock {\em The Journal of experimental biology}, 215(17):3028--3036, 2012.

\bibitem{deneubourg1991dynamics}
Jean-Louis Deneubourg, Simon Goss, Nigel Franks, Ana Sendova-Franks, Claire
  Detrain, and Laeticia Chr{\'e}tien.
\newblock The dynamics of collective sorting robot-like ants and ant-like
  robots.
\newblock In {\em Proceedings of the first international conference on
  simulation of adaptive behavior on From animals to animats}, pages 356--363,
  1991.

\bibitem{dickinson1984feeding}
Michael~H Dickinson and Charles~M Lent.
\newblock Feeding behavior of the medicinal leech, hirudo medicinalis l.
\newblock {\em Journal of Comparative Physiology A}, 154(4):449--455, 1984.

\bibitem{duives2013state}
Dorine~C Duives, Winnie Daamen, and Serge~P Hoogendoorn.
\newblock State-of-the-art crowd motion simulation models.
\newblock {\em Transportation research part C: emerging technologies},
  37:193--209, 2013.

\bibitem{friesen1993mechanisms}
WO~Friesen and RA~Pearce.
\newblock Mechanisms of intersegmental coordination in leech locomotion.
\newblock In {\em Seminars in Neuroscience}, volume~5, pages 41--47. Elsevier,
  1993.

\bibitem{garcia2005statistics}
Elizabeth Garcia-Perez, Alberto Mazzoni, Davide Zoccolan, Hugh~PC Robinson, and
  Vincent Torre.
\newblock Statistics of decision making in the leech.
\newblock {\em The Journal of neuroscience}, 25(10):2597--2608, 2005.

\bibitem{gaudry2010feeding}
Quentin Gaudry and William~B Kristan.
\newblock Feeding-mediated distention inhibits swimming in the medicinal leech.
\newblock {\em The Journal of Neuroscience}, 30(29):9753--9761, 2010.

\bibitem{Georgoudas06}
Ioakeim~G. Georgoudas, Georgios~Ch. Sirakoulis, and Ioannis~Th. Andreadis.
\newblock A simulation tool for modelling pedestrian dynamics during evacuation
  of large areas.
\newblock In {\em Artificial Intelligence Applications and Innovations}, pages
  618--626. Springer, 2006.

\bibitem{gerry2012serotonin}
Shannon~P Gerry, Amanda~J Daigle, Kara~L Feilich, Jessica Liao, Azzara~L Oston,
  and David~J Ellerby.
\newblock Serotonin as an integrator of leech behavior and muscle mechanical
  performance.
\newblock {\em Zoology}, 115(4):255--260, 2012.

\bibitem{gunji2011adaptive}
Yukio-Pegio Gunji, Tomohiro Shirakawa, Takayuki Niizato, Masaki Yamachiyo, and
  Iori Tani.
\newblock An adaptive and robust biological network based on the
  vacant-particle transportation model.
\newblock {\em Journal of theoretical biology}, 272(1):187--200, 2011.

\bibitem{heigeas2010physically}
Laure He{\"\i}geas, Annie Luciani, Joelle Thollot, and Nicolas Castagn{\'e}.
\newblock A physically-based particle model of emergent crowd behaviors.
\newblock {\em arXiv preprint arXiv:1005.4405}, 2010.

\bibitem{Helbing00}
Dirk Helbing, Illes Farkas, and Tamas Vicsek.
\newblock Simulating dynamical features of escape panic.
\newblock {\em Nature}, 407(6803):487--490, 2000.

\bibitem{Helbing02}
Dirk Helbing, Illes~J Farkas, Peter Molnar, and Tam{\'a}s Vicsek.
\newblock Simulation of pedestrian crowds in normal and evacuation situations.
\newblock {\em Pedestrian and evacuation dynamics}, 21:21--58, 2002.

\bibitem{hughes2003flow}
Roger~L Hughes.
\newblock The flow of human crowds.
\newblock {\em Annual review of fluid mechanics}, 35(1):169--182, 2003.

\bibitem{KristanJr1977191}
William B.~Kristan Jr. and Peter~B. Guthrie.
\newblock Acquisition of swimming behavior in chronically isolated single
  segments of the leech.
\newblock {\em Brain Research}, 131(1):191 -- 195, 1977.

\bibitem{kalogeiton2014hey}
Vicky~S Kalogeiton, Dim~P Papadopoulous, George~Ch Sirakoulis, Alejandro
  Vazquez-Otero, Jan Faigl, Natividad Duro, Raquel Dormido, Thomas~C Henderson,
  Anshul Joshi, Kirril Rashkeev, et~al.
\newblock Hey physarum! can you perform slam?
\newblock {\em International Journal of Unconventional Computing}, 10(4), 2014.

\bibitem{kalogeiton2015cellular}
VS~Kalogeiton, DP~Papadopoulos, IP~Georgilas, G~Ch Sirakoulis, and
  AI~Adamatzky.
\newblock Cellular automaton model of crowd evacuation inspired by slime mould.
\newblock {\em International Journal of General Systems}, 44(3):354--391, 2015.

\bibitem{kristan2005neuronal}
William~B Kristan~Jr, Ronald~L Calabrese, and W~Otto Friesen.
\newblock Neuronal control of leech behavior.
\newblock {\em Progress in neurobiology}, 76(5):279--327, 2005.

\bibitem{kristan2000development}
William~B Kristan~Jr, F~James Eisenhart, Lisa~A Johnson, and Kathleen~A French.
\newblock Development of neuronal circuits and behaviors in the medicinal
  leech.
\newblock {\em Brain research bulletin}, 53(5):561--570, 2000.

\bibitem{langridge2008experienced}
Elizabeth~A Langridge, Ana~B Sendova-Franks, and Nigel~R Franks.
\newblock How experienced individuals contribute to an improvement in
  collective performance in ants.
\newblock {\em Behavioral Ecology and Sociobiology}, 62(3):447--456, 2008.

\bibitem{Lee07}
Kang~Hoon Lee, Myung~Geol Choi, Qyoun Hong, and Jehee Lee.
\newblock Group behavior from video: a data-driven approach to crowd
  simulation.
\newblock In {\em Proceedings of the 2007 ACM SIGGRAPH/Eurographics symposium
  on Computer animation}, pages 109--118. Eurographics Association, 2007.

\bibitem{Lerner07}
Alon Lerner, Yiorgos Chrysanthou, and Dani Lischinski.
\newblock Crowds by example.
\newblock In {\em Computer Graphics Forum}, volume~26, pages 655--664. Wiley
  Online Library, 2007.

\bibitem{li2008lattice}
Xiaomeng Li, Tao Chen, Lili Pan, Shifei Shen, and Hongyong Yuan.
\newblock Lattice gas simulation and experiment study of evacuation dynamics.
\newblock {\em Physica A: Statistical Mechanics and its Applications},
  387(22):5457--5465, 2008.

\bibitem{liang2012cma}
Yu~Liang, William Melvin, Subramania~I Sritharan, Shane Fernandes, and Darrell
  Barker.
\newblock Cma-ht: a crowd motion analysis framework based on heat-transfer
  analog model.
\newblock In {\em SPIE Defense, Security, and Sensing}, pages 84020J--84020J.
  International Society for Optics and Photonics, 2012.

\bibitem{lockery1993computational}
Shawn~R Lockery and Terrence~J Sejnowski.
\newblock The computational leech.
\newblock {\em Trends in neurosciences}, 16(7):283--290, 1993.

\bibitem{lockery1993lower}
SR~Lockery and TJ~Sejnowski.
\newblock A lower bound on the detectability of nonassociative learning in the
  local bending reflex of the medicinal leech.
\newblock {\em Behavioral and neural biology}, 59(3):208--224, 1993.

\bibitem{Nishinari06}
Katsuhiro Nishinari, Ken Sugawara, Toshiya Kazama, Andreas Schadschneider, and
  Debashish Chowdhury.
\newblock Modelling of self-driven particles: Foraging ants and pedestrians.
\newblock {\em Physica A: Statistical Mechanics and its Applications},
  372(1):132--141, 2006.

\bibitem{Omer1994}
Haim Omer and Nahman Alon.
\newblock The continuity principle: A unified approach to disaster and trauma.
\newblock {\em American Journal of Community Psychology}, 22(2):273--287, 1994.

\bibitem{pearce1988model}
RA~Pearce and WO~Friesen.
\newblock A model for intersegmental coordination in the leech nerve cord.
\newblock {\em Biological cybernetics}, 58(5):301--311, 1988.

\bibitem{reynolds2006big}
Craig Reynolds.
\newblock Big fast crowds on ps3.
\newblock In {\em Proceedings of the 2006 ACM SIGGRAPH symposium on
  Videogames}, pages 113--121. ACM, 2006.

\bibitem{Shiwakoti2013}
Nirajan Shiwakoti and Majid Sarvi.
\newblock Enhancing the panic escape of crowd through architectural design.
\newblock {\em Transportation Research Part C: Emerging Technologies},
  37(0):260 -- 267, 2013.

\bibitem{Shiwakoti2011}
Nirajan Shiwakoti, Majid Sarvi, Geoff Rose, and Martin Burd.
\newblock Animal dynamics based approach for modeling pedestrian crowd egress
  under panic conditions.
\newblock {\em Transportation Research Part B: Methodological}, 45(9):1433 --
  1449, 2011.
\newblock Select Papers from the 19th \{ISTTT\}.

\bibitem{Sirakoulisbook}
Georgios~Ch. Sirakoulis and Stefania Bandini, editors.
\newblock {\em Cellular Automata: 10th International Conference on Cellular
  Automata for Research and Industry, ACRI 2012, Santorini Island, Greece,
  September 24-27, 2012. Proceedings}, volume 7495 of {\em Lecture Notes in
  Computer Science}. Springer, 2012.

\bibitem{su2014crowd}
Zhou Su, Hua Wei, and Sha Wei.
\newblock Crowd event perception based on spatiotemporal weber field.
\newblock {\em Journal of Electrical and Computer Engineering}, 2014:1, 2014.

\bibitem{taneja2011national}
Pankaj Taneja and John Rowson.
\newblock National survey of the use and application of leeches in oral and
  maxillofacial surgery in the united kingdom.
\newblock {\em British Journal of Oral and Maxillofacial Surgery},
  49(6):438--441, 2011.

\bibitem{taylor2000model}
Adam Taylor, Garrison~W Cottrell, and William~B Kristan~Jr.
\newblock A model of the leech segmental swim central pattern generator.
\newblock {\em Neurocomputing}, 32:573--584, 2000.

\bibitem{tsiftsisreal}
Anastasios Tsiftsis, Ioakeim~G Georgoudas, and Georgios~Ch Sirakoulis.
\newblock Real data evaluation of a crowd supervising system for stadium
  evacuation and its hardware implementation.

\bibitem{Vizzari13}
Giuseppe Vizzari, Lorenza Manenti, and Luca Crociani.
\newblock Adaptive pedestrian behaviour for the preservation of group cohesion.
\newblock {\em Complex Adaptive Systems Modeling}, 1(1):1--29, 2013.

\bibitem{neumann66}
John Von~Neumann, Arthur~Walter Burks, et~al.
\newblock {\em Theory of self-reproducing automata}.
\newblock University of Illinois press Urbana, 1966.

\bibitem{yang2008body}
Qinghai Yang, Junzhi Yu, Rui Ding, and Min Tan.
\newblock Body-deformation steering approach to guide a multi-mode amphibious
  robot on land.
\newblock In {\em Intelligent Robotics and Applications}, pages 1021--1030.
  Springer, 2008.

\bibitem{yang2007preliminary}
Qinghai Yang, Junzhi Yu, Min Tan, and Weibing Wang.
\newblock Preliminary development of a biomimetic amphibious robot capable of
  multi-mode motion.
\newblock In {\em Robotics and Biomimetics, 2007. ROBIO 2007. IEEE
  International Conference on}, pages 769--774. IEEE, 2007.

\bibitem{yu2009amphibious}
Shumei Yu, Shugen Ma, Bin Li, and Yuechao Wang.
\newblock An amphibious snake-like robot: design and motion experiments on
  ground and in water.
\newblock In {\em Information and Automation, 2009. ICIA'09. International
  Conference on}, pages 500--505. IEEE, 2009.

\bibitem{Yu07}
YF~Yu and WG~Song.
\newblock Cellular automaton simulation of pedestrian counter flow considering
  the surrounding environment.
\newblock {\em Physical Review E}, 75(4):046112, 2007.

\bibitem{zaccardi2004sensitization}
Maria~Luisa Zaccardi, Giovanna Traina, Enrico Cataldo, and Marcello Brunelli.
\newblock Sensitization and dishabituation of swim induction in the leech< i>
  hirudo medicinalis</i>: role of serotonin and cyclic amp.
\newblock {\em Behavioural brain research}, 153(2):317--326, 2004.

\bibitem{zheng2007systems}
Min Zheng, W~Otto Friesen, and Tetsuya Iwasaki.
\newblock Systems-level modeling of neuronal circuits for leech swimming.
\newblock {\em Journal of computational neuroscience}, 22(1):21--38, 2007.

\end{thebibliography}

\end{document}